\documentclass[preprint,12pt]{elsarticle}
\usepackage[psamsfonts]{amssymb}
\usepackage{amsmath}
\journal{}

\newcommand{\be}{\begin{equation}}
\newcommand{\ee}{\end{equation}}
\newcommand{\bea}{\begin{eqnarray}}
\newcommand{\eea}{\end{eqnarray}}

\newcommand{\n}{\nonumber} 

\begin{document}

\begin{frontmatter}
\title{Coherent states on a circle: the Higgs-like approach}

\author{Ali Mahdifar$^{1,2} $ }

\ead{a.mahdifar@sci.ui.ac.ir}

\author{Ehsan Amooghorban$^{3,4}$ }

\ead{Ehsan.amooghorban@sku.ac.ir}

\address{ $^{1}$  Department of Physics, University of Isfahan, Hezar Jerib, Isfahan, 81746-73441, Iran.}
\address{ $^{2}$ Department of Physics, Quantum optics group, University of Isfahan, Hezar Jerib St. Isfahan 81764-73441, Iran. }
\address{ $^{3}$ Department of Physics, Faculty of Science, Shahrekord University, Shahrekord, 88186-34141, Iran.}
\address{ $^{4}$ Nanotechnology Research Center, Shahrekord University, Shahrekord 88186-34141, Iran.}
\begin{abstract}
In this paper, the Higgs-like approach is used to analyze the quantum dynamics of a harmonic oscillator constrained on a circle. We obtain the Hamiltonian of this system as a function of the Cartesian coordinate of the tangent line through the gnomonic projection and then quantize it in the standard way. We then recast the Hamiltonian in a shape-invariant form and derive the spectrum energy of the confined harmonic oscillator on the circle. With help of the f-deformed oscillator algebra, we construct the coherent states on the circle and investigate their quantum statistical properties. We find that such states show nonclassical features like squeezing and sub-Poissonian statistics even in small curvatures of the circle.\\
\end{abstract}

\end{frontmatter}

\section{Introduction}
One of the fundamental problems in quantum mechanics is to find a compatible quantization procedure of a classical system. The canonical quantization is the usual well-known approach which is based on using the classical form of observables as a real function in phase space and replacing the coordinates $x$ and $p$ with operators $\hat{q}$ and $\hat{p}$ in a symmetrical order. However, this procedure works only for systems with phase space $R^{2N}$, provided that the quantization is carried out in Cartesian coordinates.
In a non-Cartesian coordinate, we can start from the Lagrangian formalism and obtain canonical conjugate momenta of the corresponding dynamical variables of the system to construct the Hamiltonian, and impose canonical commutation relations between them. However, it raises a host of difficulties like a mismatch between the canonical momenta and the Noether momenta~\cite{Shankar}. Instead, we can consistently quantize the system in Cartesian coordinates and make the change to curvilinear coordinates at the quantum level. In general, the canonical quantization becomes cumbersome when the classical coordinates and/or their conjugate momenta do not vary over the full interval from $ -\infty $ to $ +\infty $.

There are various reasons why physicists try to develop the correct quantum mechanics in curved spaces. Despite extending our knowledge on quantum mechanics in more general mathematical frameworks, it provides the capacity for application in many different fields of physics, such as the quantum Hall effect~\cite{Bracken2007}, the quantum dots~\cite{Gritsev2001,Bulaev2004}, and the coherent state quantization~\cite{Gazeau2004}. To the best of our knowledge, there are three known approaches to study nonrelativistic quantum mechanics in the spaces of constant curvature: The first one is the Noether quantization which is concerned with the identification of the Killing vector fields and the Noether momenta. In this manner, one can construct the Hamiltonian of the system as a function of Noether momenta, and then apply the quantization process to the components of these momenta \cite{Carinena1, Carinena2, Bracken}.
The second method is the so-called thin layer quantization \cite{Jensen, da Costa1, da Costa2, Dehdashti}.
In this approach, the 2D curved surface is embedded into the larger 3D Euclidean space, and the introduction of effective potential results in a dimensional reduction in the Schrodinger equation. In the last approach, which is known as the Higgs method, dynamical symmetries of a system with a spherical geometry are extracted\cite{Higgs, Leemon}. In this way, the Hamiltonian of the system is expressed in terms of the Casimir operators of the Lie algebra of such dynamical symmetries, and subsequently, the energy eigenvalues are obtained through the eigenvalues of the Casimir operators. In this contribution, we follow the Higgs-like approach.

On the other hand, the standard coherent states (CSs)~\cite{Glauber} as well as the generalized CSs associated with various algebras~\cite{Perelomov, Ali, Klauder} play an important role in different fields of physics. Among them, the so-called nonlinear or f-deformed CSs~\cite{Solomon, Manko} have attracted much interest due to their nonclassical properties, such as quantum interference and amplitude squeezing~\cite{Vogel1, Vogel2}. Such states indeed are the eigenstates of the annihilation operator of the f-deformed oscillator~\cite{Solomon, Manko} can be realized in the center of mass motion of a laser-driven trapped ion \cite{Vogel2, MahdifarVogel}, and in a micromaser under intensity-dependent atom-field interaction~\cite{Naderi}.
It is well known, the resolution of identity is the most important mathematical feature of the CSs, i.e., they form an overcomplete set. Furthermore, the CSs include properties of the space that the relevant system defined on it. Therefore, if we consider, for example, the harmonic oscillator on a curved space and construct the associated CSs, they contain the curvature effects. Thus, employing the curvature-dependent CSs manifold makes it possible to investigate the curvature effects on the physical phenomena within the framework of the CSs approach \cite{IJGMMP}. This can enable us to study the curvature effects of some physical structures like nanostructures on their electronic and optical properties  \cite{Popov}.

In Refs. \cite{Mahdifar2006} and \cite{KC}, based on the Higgs model, the generalized (nonlinear) coherent states of a two-dimensional harmonic oscillator and a Kepler-Coulomb potential on a spherical surface are constructed and some of their quantum optical properties are studied. Later on, the thermal nonlinear coherent states on a sphere is presented in \cite{Bagheri} and the relation of the nonclassical properties of the constructed CSs with the curvature and temperature is investigated.

We have recently presented the classical and quantum-mechanical treatment of the Higgs model in the presence of dissipation using a continuum of oscillators as a reservoir~\cite{Amooghorban}. There, we showed that the transition probability between energy levels is dependent on the curvature of the physical space and the dissipative effect, and significant probabilities of transition are only possible if the transition and reservoir's oscillators frequencies are almost in resonance.
We have then investigated the dynamics of both a free particle and an isotropic harmonic oscillator constrained to move on a spheroidal surface using two consecutive gnomonic projections~\cite{Spheroid}. We have obtained the Hamiltonian of the systems as a function of the Cartesian coordinates of the tangent plane and finally quantized them. We have shown that the effect of nonsphericity of the surface appears like an effective potential. We found that the deviation from the sphericity leads to the split of the energy levels of the constrained oscillator on a sphere and lifts the degeneracy.

With the above background, the goal of this contribution is twofold. In the first step, we intend to obtain the classical and quantum Hamiltonian of a harmonic oscillator confined on a circle which is viewed as the simplest curvature in geometry. To do this, we use the gnomonic projection, which is a projection onto the tangent line from the center of the circle into the embedding line. We first obtain the metric of the circle background in the gnomonic coordinates and construct the Hamiltonian of the harmonic oscillator. We follow the canonical quantization method and derive the quantum counterpart of the classical Hamiltonian.
We then recast the Hamiltonian in a shape-invariant form and compute the exact spectrum energy of the confined harmonic oscillator on a circle. We show that in the limit of the large radius of the circle (straight line), it properly reduces to the well-known energy spectrum of the conventional quantum harmonic oscillator. In the second step, we obtain a deformation function for such oscillator utilizing the f-deformed oscillator algebra. In the end, we construct the CSs on a circle employing the nonlinear CSs approach and investigate their quantum statistical properties.

This paper is organized as follows: In Sec.~\ref{QHO}, by making use of the gnomonic projection onto the tangent line, we obtain the Hamiltonian of a harmonic oscillator confined on a circle. We subsequently quantize the aforementioned Hamiltonian using a factorization method and calculate its exact spectrum energy. In Sec.~\ref{f deformed}, we obtain a deformation function for the oscillator on a circle. In Sec.~\ref{NCSOC}, we construct the corresponding CSs and examine their resolution of identity. The Sec.~ \ref{QS NCSOC} is devoted to the study of the quantum statistical properties of the constructed CSs. Especially, the influence of curvature of the circle on their nonclassical properties is analyzed. Finally, the summary and concluding remarks are given in section \ref{SUM}.

\section{Quantum harmonic oscillator on circle}\label{QHO}

In this section, we obtain the Hamiltonian of a harmonic oscillator confined on a circle. To this end, we obtain the metric of the circle background in the gnomonic coordinates and then construct the Hamiltonian of the harmonic oscillator.

Let us designate the Cartesian coordinates of the circle background by $(q_{1},q_{2})$, and assume that they satisfy the circle equation,
 \be\label{circle}
   q_{1}^{2}+q_{2}^{2}=R^{2},
 \ee
where, $R$ is the radius of the circle.
\begin{figure}[!ht]
\centering
\includegraphics[scale=0.3]{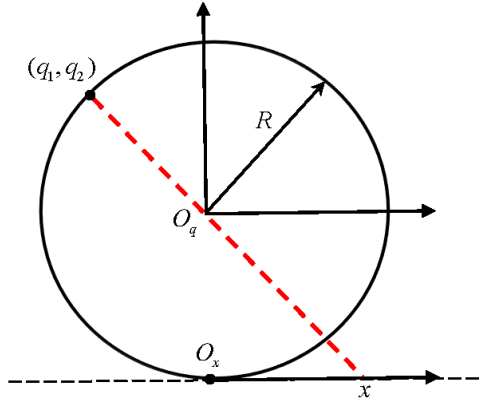}
\caption{Coordinate systems and projection from a circle with radius $R$ onto a line. The gnomonic projection of the point from
the circle's center onto the point $x$ on the tangent line.}
\label{fig1}
\end{figure}
If we denote the Cartesian coordinate of the tangent line to the
circle by $x$ (see Fig.~\ref{fig1}), the relationship between these two coordinates is given by
 \bea\label{gno}
   q_{1}&=&\frac{x}{\sqrt{[1+\lambda x^{2}}]},\n\\
   q_{2}&=&\frac{1}{\sqrt{\lambda[1+\lambda x^{2}}]},
 \eea
where, $\lambda=\frac{1}{R^{2}}$ is the
curvature of the circle. Accordingly, a point
on the circle can be represented as,
 \be
   \vec{r}\equiv(q_{1},q_{2})=(\frac{x}{\Lambda},\frac{1}{\sqrt{\lambda}\Lambda}),
 \ee
where,
 \be
   \Lambda=\sqrt{1+\lambda x^{2}}.
 \ee
Now the differential of $\vec{r}$ is given by,
 \be
   d \vec{r}=\vec{r}_{x} dx=(\frac{1}{\Lambda^{3}},-\frac{\sqrt{\lambda}x}{\Lambda^{3}}) dx.
 \ee
Here, $\vec{r}_{x}$ is the
derivatives of $\vec{r}$ with respect to $x$. After a straightforward calculation, we obtain the
metric of the circle as,
 \bea\label{ds}
   ds^{2}
   &\doteq&
   d\vec{r}\cdot d\vec{r}=(\frac{1}{\Lambda^{6}}+\frac{\lambda x^{2}}{\Lambda^{6}}) dx^{2}
                         =\frac{1}{\Lambda^{4}}dx^{2}.
 \eea
From equation (\ref{ds}), we get
 \bea
   \dot{s}^{2}=\frac{\dot{x}^{2}}{\Lambda^{4}}.
  \eea
We can therefore write the Lagrangian of a harmonic oscillator on circular background as,
 \bea\label{Lag}
   L
   &\doteq&
   \frac{1}{2}\dot{s}^{2}-V_{\rm HO}(x)=\frac{1}{2}\frac{\dot{x}^{2}}{\Lambda^{4}}-V_{\rm HO}(x),
 \eea
where $V_{\rm HO}(x)$ is the potential of the confined oscillator on the circle in terms of the tangent coordinate $x$. In what follows, we obtain the aforementioned potential. Let $s$ be the arc-length of the geodesics from the north pole of the circle, ${\vec r}_{0}=(0,R)$, to the point ${\vec r}=(q_{1}, q_{2})$. Using Eq.(\ref{circle}), we have
\be\label{the square of the distance}
  s^2=q_{1}^{2}+q_{2}^{2}=g\ q_{1}^{2}.
\ee
Here, the one-dimensional metric tensor on the circle is given by
\begin{eqnarray}\label{metric tensor on spheroid}
 g=1+\frac{q_{1}^{2}}{R^{2}-q_{1}^{2}}.
\end{eqnarray}
Accordingly, the potential energy for a harmonic oscillator on the circular curve can be written as (in this paper we put $ \hbar=m=\omega= 1)$:
 \be\label{v}
  V_{\rm HO}(s)\equiv\frac{1}{2}s^{2}=\frac{1}{2}\ g\ q_{1}^{2}
  =(1+\frac{q_{1}^{2}}{R^{2}-q_{1}^{2}})q_{1}^{2}.
 \ee

Let us reconsider the gnomonic projection, which is the projection onto the tangent line from the center of the circle. By making use of Eq.~(\ref{gno}), the points on the circle are expressed in terms of the coordinates of this projection. Thus, the potential of the oscillator on the circle in terms of $x$ is given by:
 \be\label{po}
  V_{\rm HO}(x)=\frac{1}{2}(1+\lambda x^{2})(\frac{x^2}{1+\lambda x^{2}})=\frac{1}{2}x^{2}.
 \ee
Now, by using the Lagrangian (\ref{Lag}) and the potential (\ref{po}), we can calculate the momentum as,
 \bea
   p
   =   \frac{\partial L}{\partial \dot{x}}
   =   \frac{\dot{x}}{\Lambda^{4}}.
 \eea
With proper calculation, we get  the classical Hamiltonian of the harmonic
oscillator constrained to the circular background as,
 \be\label{hclass}
   H_{\rm cl}\doteq
   \dot{x}p-L=\frac{1}{2}\left[(1+\lambda x^{2})p\right]^{2}+\frac{1}{2}x^{2}
             \equiv \frac{1}{2}\pi^{2}+\frac{1}{2}x^{2},
 \ee
where $\pi=(1+\lambda x^{2})p$. By replacing the classical momentum $p$ by related operator $(-i\frac{d}{dx})$, we can obtain the quantum counterpart of the classical Hamiltonian (\ref{hclass}) as,
 \bea\label{hquantum}
          H&=&\frac{1}{2}\left[(1+\lambda x^{2})(-i\frac{d}{dx})(1+\lambda x^{2})(-i\frac{d}{dx})\right]+\frac{1}{2}x^{2}\n\\
          &=&\frac{1}{2}\left[-(1+\lambda x^{2})^{2}\frac{d^{2}}{dx^{2}}-2\lambda x(1+\lambda x^{2})\frac{d}{dx}\right]+\frac{1}{2}x^{2}.
 \eea
Now, for any real number $\gamma$ we define the following linear operator:
 \be
   A(\gamma)=\frac{1}{\sqrt{2}}[\imath \pi+\gamma x]=\frac{1}{\sqrt{2}} \left[(1+\lambda x^{2})\frac{d}{dx}+\gamma x\right].
 \ee
Indeed, the adjoint of $A(\gamma)$ is given by:
 \be
   A^{\dag}(\gamma)=\frac{1}{\sqrt{2}}[-\imath \pi+\gamma x]=\frac{1}{\sqrt{2}} \left[-(1+\lambda x^{2})\frac{d}{dx}+\gamma x\right].
 \ee
We can easily show that
 \bea\label{fact}
   \tilde{H}_{1}(\gamma)
   &:=&
   A^{\dag}(\gamma)A(\gamma)\\
   &=&
   \frac{1}{2}\left[-(1+\lambda x^{2})^{2}\frac{d^{2}}{dx^{2}}-2\lambda x(1+\lambda x^{2})\frac{d}{dx}-\gamma+(\gamma^{2}-\lambda\gamma)x^{2}\right].\n
 \eea
By choosing $\gamma=\frac{\lambda+\sqrt{\lambda^{2}+4}}{2}\equiv\tilde{\gamma}$, for which $(\gamma^{2}-\lambda\gamma)=1$, we can get the following factorization form:
 \be\label{e1}
   \tilde{H}_{1}=A^{\dag}A=H-\frac{\gamma}{2},
 \ee
where $H$ is the Hamiltonian of the quantum oscillator~(\ref{hquantum}).
On the other hand, for the partner Hamiltonian $\tilde{H}_{2}:= AA^{\dag}$ we find
 \bea\label{Pfact}
   \tilde{H}_{2}(\gamma)
   &=&
   A(\gamma)A^{\dag}(\gamma)\\
   &=&
   \frac{1}{2}\left[-(1+\lambda x^{2})^{2}\frac{d^{2}}{dx^{2}}-2\lambda x(1+\lambda x^{2})\frac{d}{dx}+\gamma+(\gamma^{2}+\lambda\gamma)x^{2}\right].\n
 \eea
Now, after straightforward calculations, the following relationship is hold:
 \bea\label{Pfact}
   \tilde{H}_{2}(\gamma)
   =\tilde{H}_{1}(\gamma_{1})+R(\gamma_{1}),
 \eea
where $\gamma_{1}=\gamma+\lambda$ and $R(\gamma)=\gamma-\frac{\lambda}{2}$.
As it is seen, the Hamiltonian $\tilde{H}_{1}(\gamma)$ admitting a factorization form $A^{\dag}(\gamma)A(\gamma)$ in such a way that the partner Hamiltonian $\tilde{H}_{2}(\gamma)=A(\gamma)A^{\dag}(\gamma)$ is of the form as $\tilde{H}_{1}(\gamma)$ but for a different value of the $\gamma$. In this case, it is usually said that there is shape invariance and makes it possible to calculate exact spectrum of the Hamiltonian \cite{SUSY, SUSY2}. The ground state $|\psi_0(\gamma)\rangle$ with zero energy is given by $A(\gamma)|\psi_0(\gamma)\rangle=0$. Subsequently, by using Eq. (\ref{Pfact}) it is seen that $|\psi_0(\gamma_1)\rangle$ is an eigenstate of $\tilde{H}_{2}(\gamma)$ with the energy $\tilde{E}_1=R(\gamma_1)$:
 \be
   \tilde{H}_{2}(\gamma)|\psi_0(\gamma_{1})\rangle=[\tilde{H}_{1}(\gamma_{1})+R(\gamma_{1})]\ |\psi_0(\gamma_{1})\rangle=R(\gamma_{1})|\psi_0(\gamma_{1})\rangle.
 \ee
Similarity, using Eq.~(\ref{fact}) we have
 \bea
   \tilde{H}_{1}(\gamma)[A^{\dag}(\gamma)|\psi_0(\gamma_1)\rangle]
   &=&
   A^{\dag}(\gamma)\tilde{H}_{2}(\gamma)|\psi_0(\gamma_1)\rangle\n\\
   &=&
   A^{\dag}(\gamma)[\tilde{H}_{1}(\gamma_{1})+R(\gamma_{1})]\ |\psi_0(\gamma_1)\rangle\n\\
   &=&
   R(\gamma_{1})[A^{\dag}(\gamma)|\psi_0(\gamma_1)\rangle].
 \eea
In other words, $A^{\dag}(\gamma)|\psi_0(\gamma_1)\rangle$ is the first excited state of  $\tilde{H}_{1}(\gamma)$ with the energy $\tilde{E}_1=R(\gamma_1)$. We can iterate this process and obtain the energy eigenvalues of $\tilde{H}_{1}(\gamma)$ as following:
 \be
   \tilde{E}_n=\sum_{k=1}^{n}R(\gamma+k\lambda)=n(\gamma+\frac{\lambda}{2}n).
 \ee
Finally, by substituting $\gamma=\tilde{\gamma}=\frac{\lambda+\sqrt{\lambda^{2}+4}}{2}$ and using Eq. (\ref{e1}), we can obtain the eigenvalues $E_n$ of the harmonic
oscillator on the circle as
 \be\label{energy}
   E_n=\tilde{E}_n+\frac{\tilde{\gamma}}{2}=\tilde{\gamma}(n+\frac{1}{2})+\frac{\lambda}{2}n^2.
 \ee
It is obvious that in the flat limit $\lambda\rightarrow 0$, $\tilde{\gamma}\rightarrow 1$ and Eq.~(\ref{energy}) reduces to the energy eigenvalues of the quantum harmonic oscillator on a flat line, i.e, $E_{n(\lambda=0)}=n+\frac{1}{2}$.

\section{Circular oscillator Hamiltonian as an $f$-deformed quantum oscillator}\label{f deformed}

The $f$-deformed quantum oscillators \cite{Manko},
as the nonlinear oscillators are
characterized by the following deformed dynamical variables
$\hat{A}$ and  $\hat{A}^\dag$
\begin{eqnarray}\label{fd}
      \hat{A}&=&\hat{a}f(\hat{n})=f(\hat{n}+1)\hat{a},\nonumber\\
   \hat{A}^\dag&=&f(\hat{n})\hat{a}^\dag=\hat{a}^\dag f(\hat{n}+1),
    \end{eqnarray}
where $\hat{a}$ and  $\hat{a}^\dag$ are usual bosonic annihilation and creation operators.
The deformation function $f(\hat{n})$ is an operator-valued function of the
number operator $\hat{n}\ (=\hat{a}^\dag\hat{a})$ and the nonlinear properties of this system are governed by it. As usual,
without loss of generality, we choose $f(n) = f^{\dag}(n)$.
From equation (\ref{fd}), it
follows that the $f$-deformed operators $\hat{A}$,
$\hat{A}^\dag$ and $\hat{n}$ satisfy the following closed
algebra
\begin{eqnarray}\label{algebrafd}
&&[\hat{A}, \hat{A}^\dag]=
(\hat{n}+1)f^2(\hat{n}+1)-\hat{n}f^2(\hat{n}),\nonumber\\
&&[\hat{n}, \hat{A}]=-\hat{A},\,\,\,\,\,\,\,\,\, [\hat{n}, \hat{A}^{\dag}]=\hat{A}^{\dag}.
\end{eqnarray}
The above-mentioned algebra represents a deformed Weyl-Heisenberg algebra whose nature depends on the nonlinear
deformation function $f(\hat{n})$. An $f$-deformed oscillator is a nonlinear  system characterized by a Hamiltonian of the harmonic oscillator form
     \begin{equation}\label{hamiltf}
       \hat{H}=\frac{1}{2}(\hat{A}^\dag\hat{A}+\hat{A}\hat{A}^\dag).
     \end{equation}
Using Eq.~(\ref{fd}) and the number state representation $\hat{n}|n \rangle=n|n \rangle$, the eigenvalues of the     Hamiltonian (\ref{hamiltf}) can be written as
     \begin{equation}\label{energyf}
     E_n=\frac{1}{2}[(n+1)f^2(n+1)+nf^2(n)].
     \end{equation}
It is worth noting that in the limiting case $f(n)\rightarrow 1$, the deformed algebra (\ref{algebrafd}) and the deformed energy eigenvalues~(\ref{energyf}), respectively, will reduce to the standard Weyl-Heisenberg algebra and the harmonic oscillator spectrum.

Comparing the energy spectrum of the harmonic oscillator on a circular background, Eq.~(\ref{energy}), with the energy spectrum of an $f$-deformed oscillator, Eq.~(\ref{energyf}), we obtain the corresponding deformation function for the oscillator on the circle as
   \begin{equation}\label{f}
     f(\hat{n})=\left[\tilde{\gamma}+(\frac{\hat{n}-1}{2})\lambda\right]^{1/2}.
   \end{equation}
Furthermore, the annihilation and creation operators of the harmonic oscillator on the circle can be written in terms of the conventional operators $\hat{a}$ and $\hat{a}^\dag$ as follows
   \begin{equation}\label{a mpt}
     \hat{A}=\hat{a}\left[\tilde{\gamma}+(\frac{\hat{n}-1}{2})\lambda\right]^{1/2},
     \quad    \quad
     \hat{A}^\dag=\left[\tilde{\gamma}+(\frac{\hat{n}-1}{2})\lambda\right]^{1/2}\hat{a}^\dag.
   \end{equation}
These two operators satisfy the deformed Weyl-Heisenberg commutation relation
   \begin{equation}\label{algebra mpt}
     [\hat{A},
     \hat{A}^\dag]=\hat{n}\lambda+\tilde{\gamma},
   \end{equation}
and act upon the quantum number states $|n\rangle$, corresponding to the energy eigenvalues (\ref{energy}), as
     \begin{eqnarray}\label{def ladder}
     \hat{A}|n\rangle&=&f(n)\sqrt{n}|n-1 \rangle, \nonumber\\
     \hat{A}^{\dag}|n\rangle&=&f(n+1)\sqrt{n+1}|n+1 \rangle.
     \end{eqnarray}
\section{Nonlinear CSs on a circle}\label{NCSOC}
We can construct CSs for the $f$-deformed oscillator similar to those of the harmonic oscillator. The nonlinear transformation of the creation and annihilation operators naturally provides the notion of nonlinear CSs or $f$-coherent states. These states as eigenstates of the $f$-deformed annihilation operator defined~\cite{Manko},
  \be
    \hat{A}|z,f\rangle= z |z,f\rangle
  \ee
From Eq.~(\ref{fd}), we can obtain an explicit form of the nonlinear coherent states on a circle (NCSsOC) in the number state representation as,
  \begin{equation}\label{alfa f}
       |z,f\rangle=\mathcal{N}(|z|^{2})^{-1/2}\sum_{n=0}^{\infty}\frac{z^n}{\sqrt{n!}f(n)!}|n\rangle,
  \end{equation}
where by definition $f(0)!=1$ and
  \be
    f(n)!=f(n)f(n-1)\cdots f(1),
  \ee
$z$ is a complex number, and the normalization
constant $\mathcal{N}$ is given by
  \be
    \mathcal{N}(|z|^{2})=\sum_{n=0}^{\infty}\frac{|z|^{2n}}{n![f(n)!]^2}.
  \ee
Here, the deformation function corresponding to the oscillator on a circle $f(n)!$, using Eq. (\ref{f}) and after some calculations, is given by
  \be
    f(n)!=(\frac{\lambda}{2})^{n} \frac{\Gamma(\beta+n)}{\Gamma(\beta)},
  \ee
where $\beta=\frac{2 \tilde{\gamma}}{\lambda}+1$ and $\Gamma$ is the gamma function. Similarly, we can define the f-deformed CSs corresponding to the harmonic oscillator on the circle as below:
\begin{equation}\label{NCSs}
       |z,f\rangle_{circle}\equiv|z\rangle_{\lambda}=\mathcal{N}(|z|^{2})^{-1/2}
       \sum_{n=0}^{\infty}\frac{{z}^{n}}{\sqrt{\rho(n)}}|n\rangle,
\end{equation}
where $\rho(n)=n!\ (\frac{\lambda}{2})^{2n}\left[\frac{\Gamma(\beta+n)}{\Gamma(\beta)}\right]^{2}$.
It is worth noting that to have states belonging to the Fock space, is required that $0 < \mathcal{N}(|z|^{2}) <\infty $, which implies that $|z|\leq \lim_{n\rightarrow \infty} n[f(n)]^2=\infty$ , where in the last equality we have used Eq. (\ref{f}).
Moreover, it is obvious that in the flat limit, $\lambda\rightarrow 0$, $\rho(n) \rightarrow n!$ and the above deformed CSs reduce to the standard CSs:
 \begin{equation}\label{CSs}
       |z\rangle=e^{-\frac{|z|^{2}}{2}}
       \sum_{n=0}^{\infty}\frac{{z}^{n}}{\sqrt{n!}}|n\rangle.
\end{equation}
\subsection{Resolution of identity}
In this subsection, let us show the constructed NCSsOC form an overcomplete set. In other words, the following resolution of identity has to be satisfied
  \begin{eqnarray}\label{25}
        \int d^2 z\ |z\rangle_{\lambda}\ W( | z |^2)\ _{\lambda}\langle z|  = \sum_{n = 0}^N  | n \rangle \langle n |= I.
 \end{eqnarray}
The resolution of identity can be achieved by finding a measure function $W(|z|^2)$. For this purpose, we
substitute $|z\rangle_{\lambda}$ from the Eq.~(\ref{NCSs}) into Eq.~(\ref{25}), and yields
\begin{eqnarray}\label{26}
       \int d^2 z\ |z\rangle_{\lambda}\ W( | z |^2)\ _{\lambda}\langle z| =
       \pi \sum_{n=0}^\infty\frac{ | n \rangle \langle n |}{\rho(n)}
       \int d(|z|^2) |z|^{2n} \frac{W(|z|^2)}{\mathcal{N}(|z|^{2})},
\end{eqnarray}
where we have used: $z = |z|{e^{i\theta }}$ and ${d^2}z = \frac{1}{2}d{(|z|^2)}d\theta $. By using the change of variable $x=|z|^{2}$ and considering $ w(x) = \pi \frac{W( | z |^2)}{\mathcal{N}(|z|^{2})}$, we have:
\begin{equation}\label{27}
      \int_0^\infty  w(x){x^n}dx =\rho(n)=\frac{1}{[(\frac{2}{\lambda})^{2}]^n} \frac{\Gamma(n+1)\ \Gamma^{2}(\beta+n)}{\Gamma^{2}(\beta)}.
\end{equation}
The above integral is called the moment problem and well-known mathematical methods such as Mellin
transformations can be used to solve it~\cite{Sixdeniers, Mathai}. From definition the of Meijers $G$-function, it follows that,
 \begin{eqnarray}\label{28}
    \int dx\ x^{k - 1} G_{p,q}^{m,n}
    \left( \alpha x \Big{|}
     \begin{array}{*{20}{c}}
      {a_1,\cdots,a_n,a_{n + 1},\cdots,a_p}\\
      {b_1,\cdots,b_m,b_{m + 1},\cdots,b_q}
     \end{array}
    \right) \n\\
    = \frac{1}{{{\alpha^k}}}\frac{{\prod _{j = 1}^m\Gamma ({b_j} + k)\prod _{j = 1}^n\Gamma (1 - {a_j} - k)}}{{\prod _{j    = m + 1}^q}\Gamma (1 - {b_j} - k){\prod _{j= n + 1}^p}\Gamma ({a_j} + k)}.
\end{eqnarray}
Comparing Eq.~(\ref{27}) with Eq.~(\ref{28}), we find that the measure function can be written as,
\begin{eqnarray}\label{29}
   w(x)
   &=&
   \frac{4}{\lambda^2\Gamma^{2}(\beta)}
   G_{0,3}^{3,0}
   \left( \frac{4x}{\lambda}^{2}\  \Big{|}
     \begin{array}{*{20}{c}}
      {0}\\
      {0,\  \beta-1,\ \beta-1}\\
     \end{array}
   \right).
\end{eqnarray}
In this manner, it is seen that the NCSsOC satisfy the resolution of identity and consequently form an overcomplete set.
\section{Quantum statistical properties of the NCSsOC}\label{QS NCSOC}
In this section, we turn to investigate some quantum optics features of the NCSsOC, such as probability of finding $n$ quanta, mean number of photons, the Mandel parameter and quadrature squeezing.
\subsection{Photons-number distribution}
\begin{figure}[t!]
\centering
\includegraphics[scale=0.45]{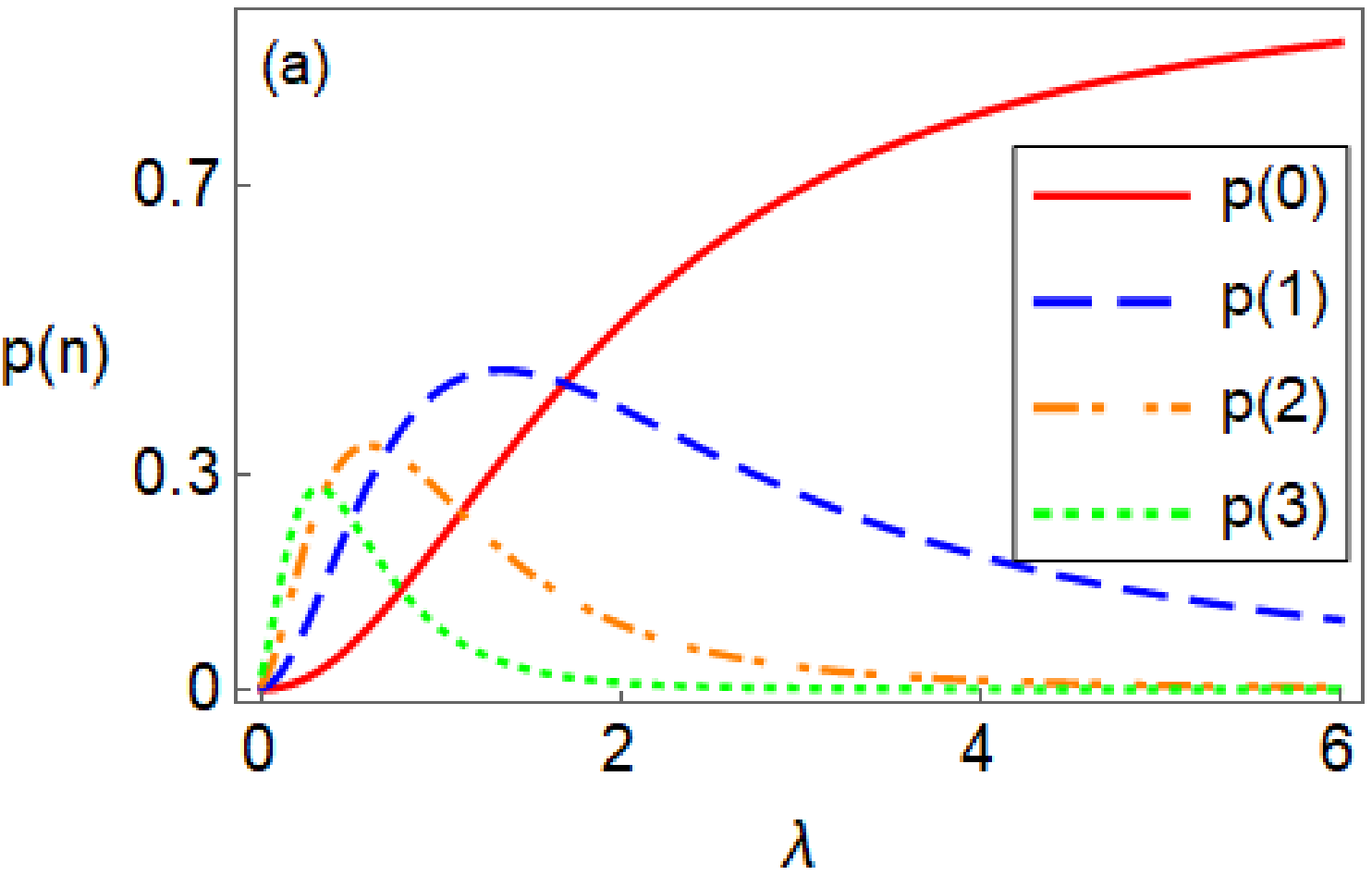}
\includegraphics[scale=0.42]{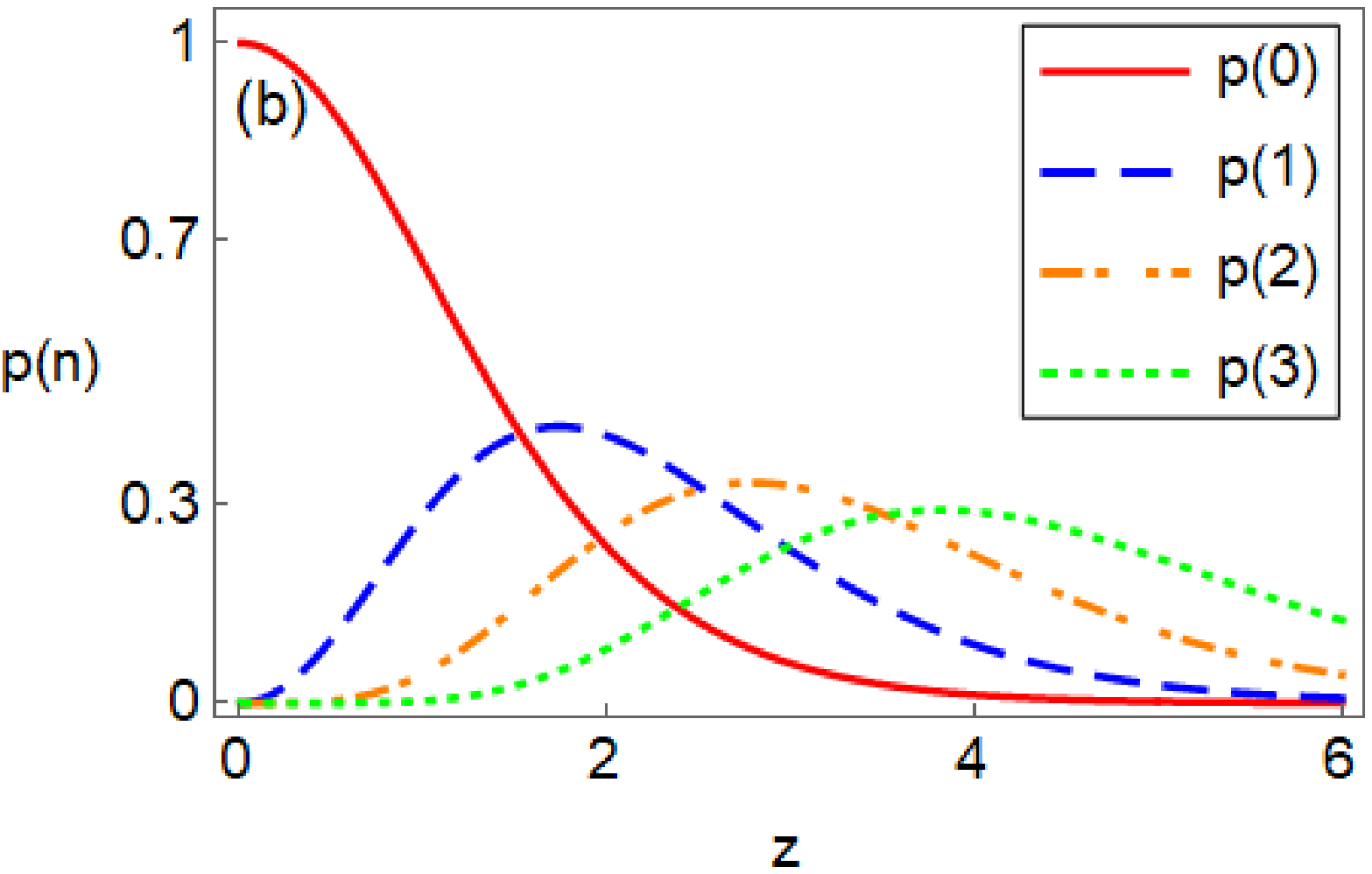}
\includegraphics[scale=0.42]{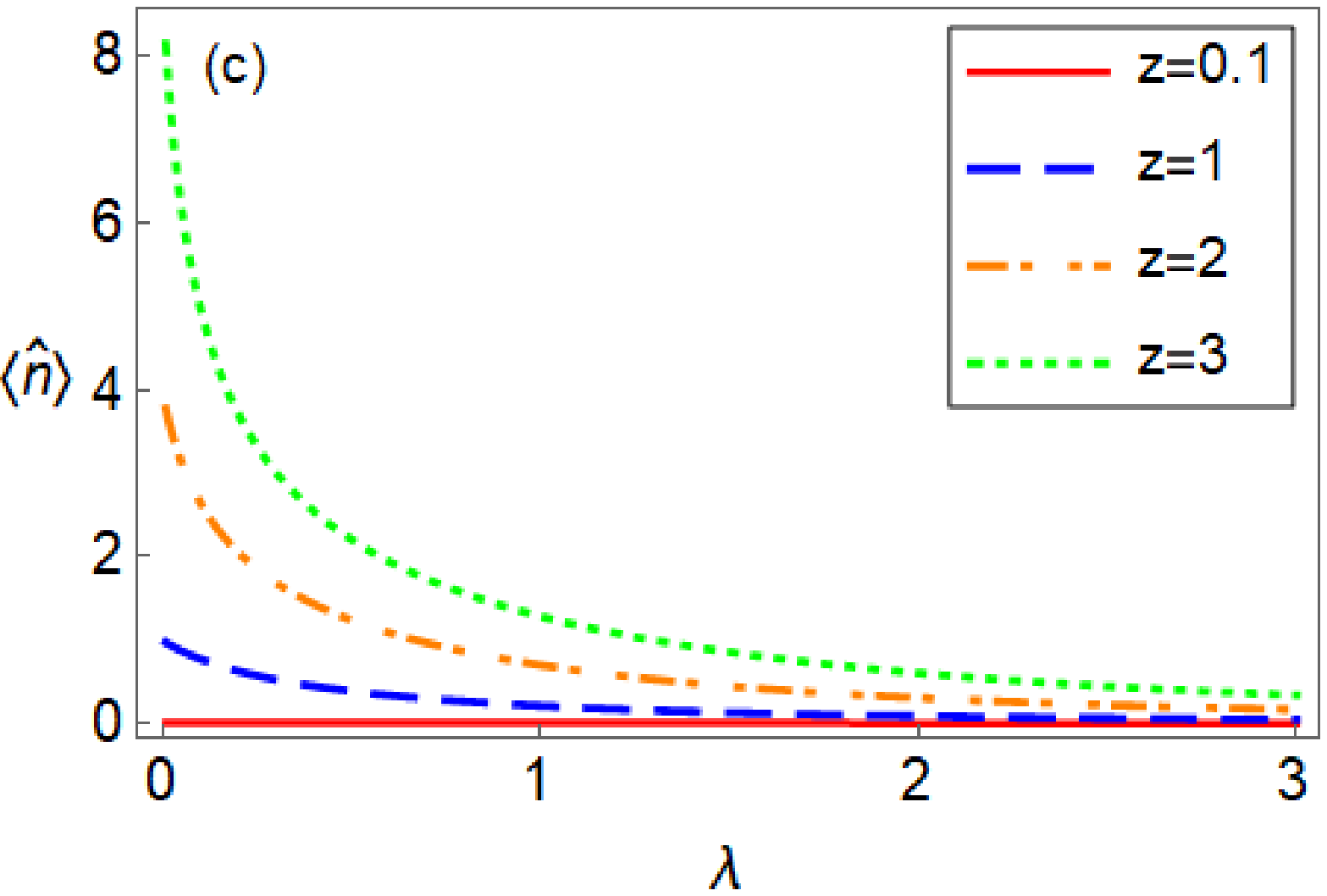}
\includegraphics[scale=0.43]{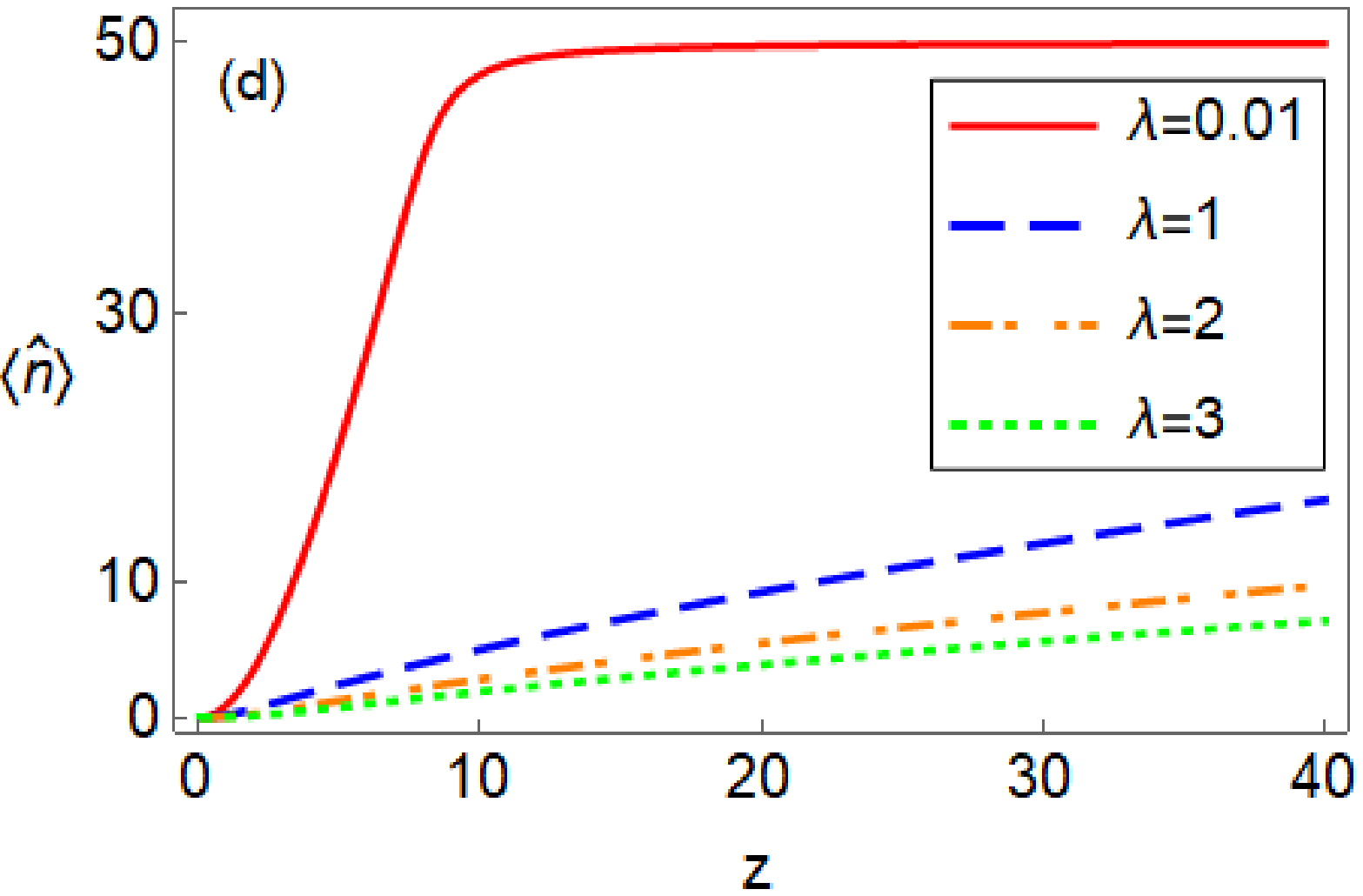}
\caption{ The probability of finding $n$ photon in the NCSsOC as a function of $\lambda$ and $z$ (a,b). Here, $z=3$ and  $\lambda=0.5$ in panels (a) and (b), respectively. The mean number of  photons in the NCSsOC versus $\lambda$ and $z$ (c,d).}
\label{fig:photons mean number}
\end{figure}

Using Eq.(27), the mean number of photons in the NCSsOC is obtained as follows:
\begin{equation}\label{mean number of photons}
\langle \hat{n} \rangle =  \ _{\lambda}\langle z| {\hat a^\dag }\hat a |z\rangle_{\lambda} =\sum_{n=0}^{N} n p(n),
\end{equation}
where the probability of finding $n$ photon in the NCSsOC is given by
\begin{equation}\label{probability of finding n photon}
p(n)=\frac{|z|^{2n}}{\mathcal{N}(|z|^2)\rho(n)}.
\end{equation}
As it is clear from the above equation, it is difficult to predict the results analytically. Thus, in Fig.~\ref{fig:photons mean number} we show the effect of the parameters $z$ and $\lambda$ on the probability of finding $n$ photons and also the mean number of photons in the NCSsOC.

In Fig.~\ref{fig:photons mean number}(a) the probability $p(n)$ of the NCSsOC for $n >1 $ is characterized by a fast initial increase followed by a very slow decrease and tends to zero, while the $p(0)$ always enhances with increasing $\lambda$ and eventually approaches unity, as expected from the conservation of probability. This behavior is also true for all other values of $z$. Moreover, the probability of finding $n$ photon for the large value of $n $ is less than the NCSsOC with small photons.
In Fig.~\ref{fig:photons mean number}(b), the probability $p(n)$ for $n >1 $ show the same trend in Fig.~\ref{fig:photons mean number}(b). However, the probability $p(0)$ decreases from its initial unit-value with increasing $z$ and eventually approaches zero.
To get further insight, let us consider the limiting case $\lambda\rightarrow 0$. From Eq.~(\ref{probability of finding n photon}), we find that the probability $p(n)$ tends to a Poissonian distribution, and the NCSsOC approaches the standard CSs.

The mean number of photons in the NCSsOC as a function of $\lambda$ and $z$ are shown in Fig.~\ref{fig:photons mean number}(c) and ~\ref{fig:photons mean number}(d), respectively. These results show that the highest value of the photons mean number occurs at the large (small) value of $z$ ($\lambda$). In other words, the mean number of photons in the NCSsOC is not significant when both parameters $z$ and $\lambda$ are simultaneously small or large.
\subsection{Mandel parameter}
\begin{figure}[t!]
\centering
\includegraphics[scale=0.48]{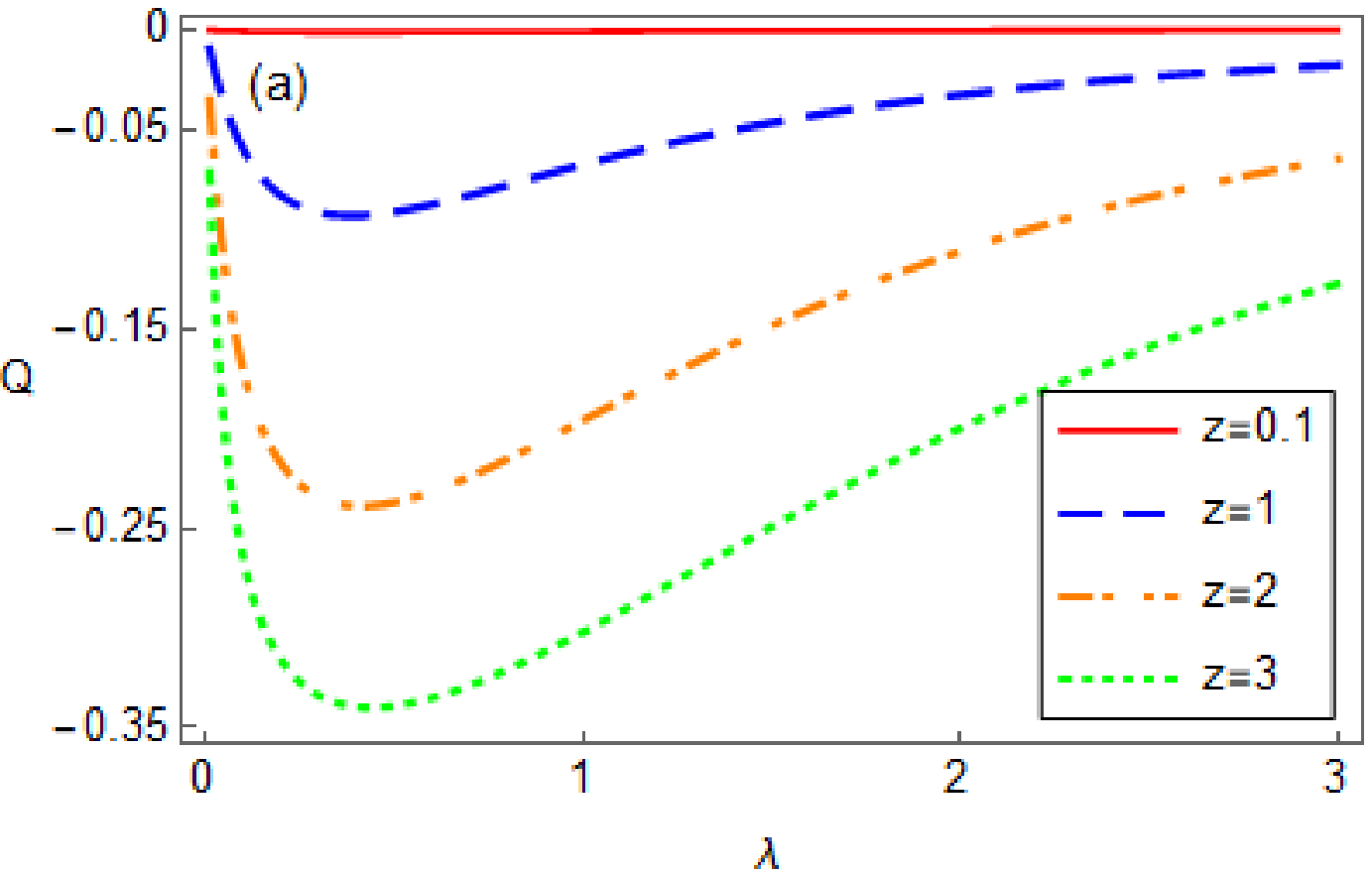}
\includegraphics[scale=0.42]{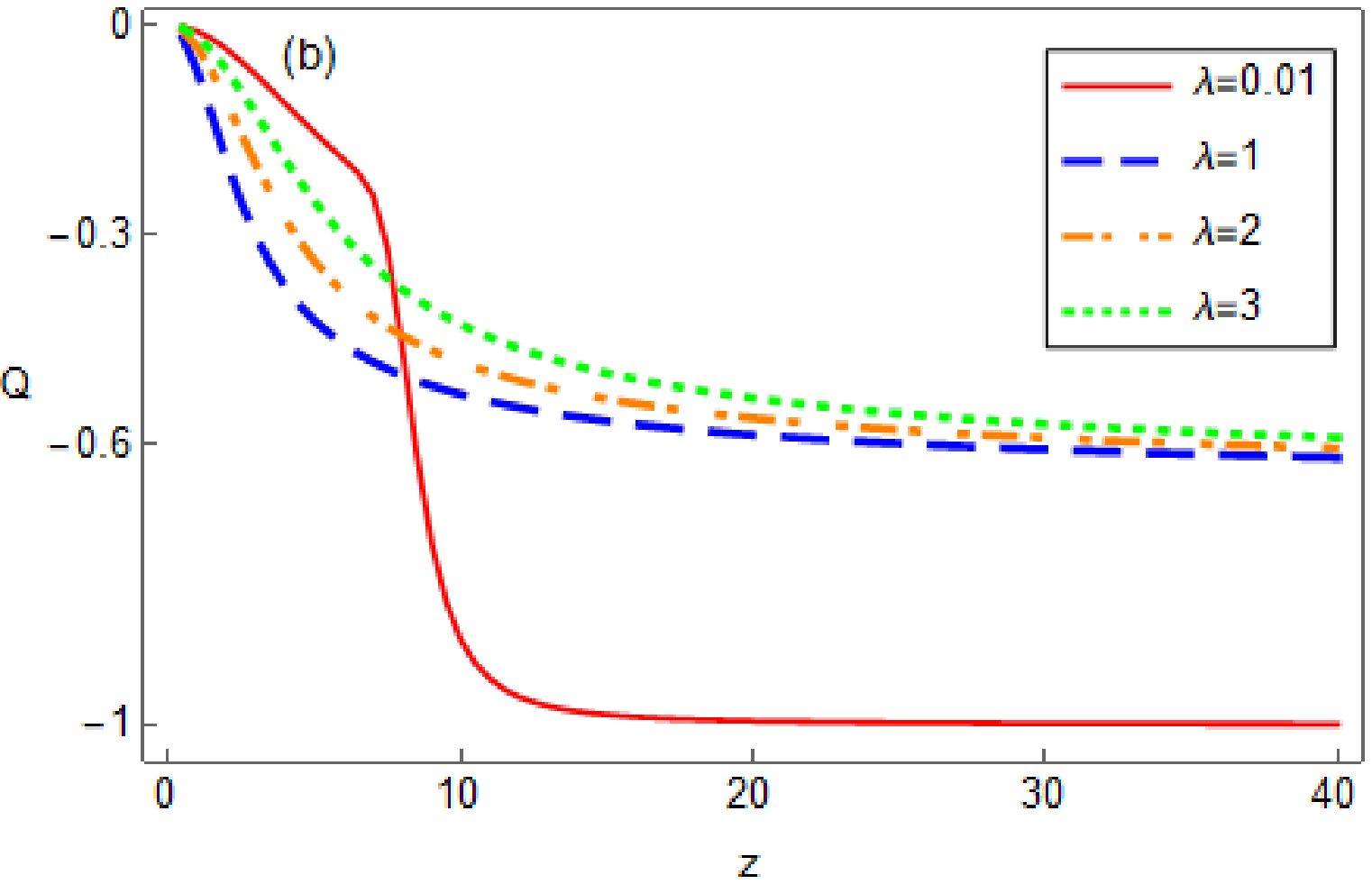}
\includegraphics[scale=0.5]{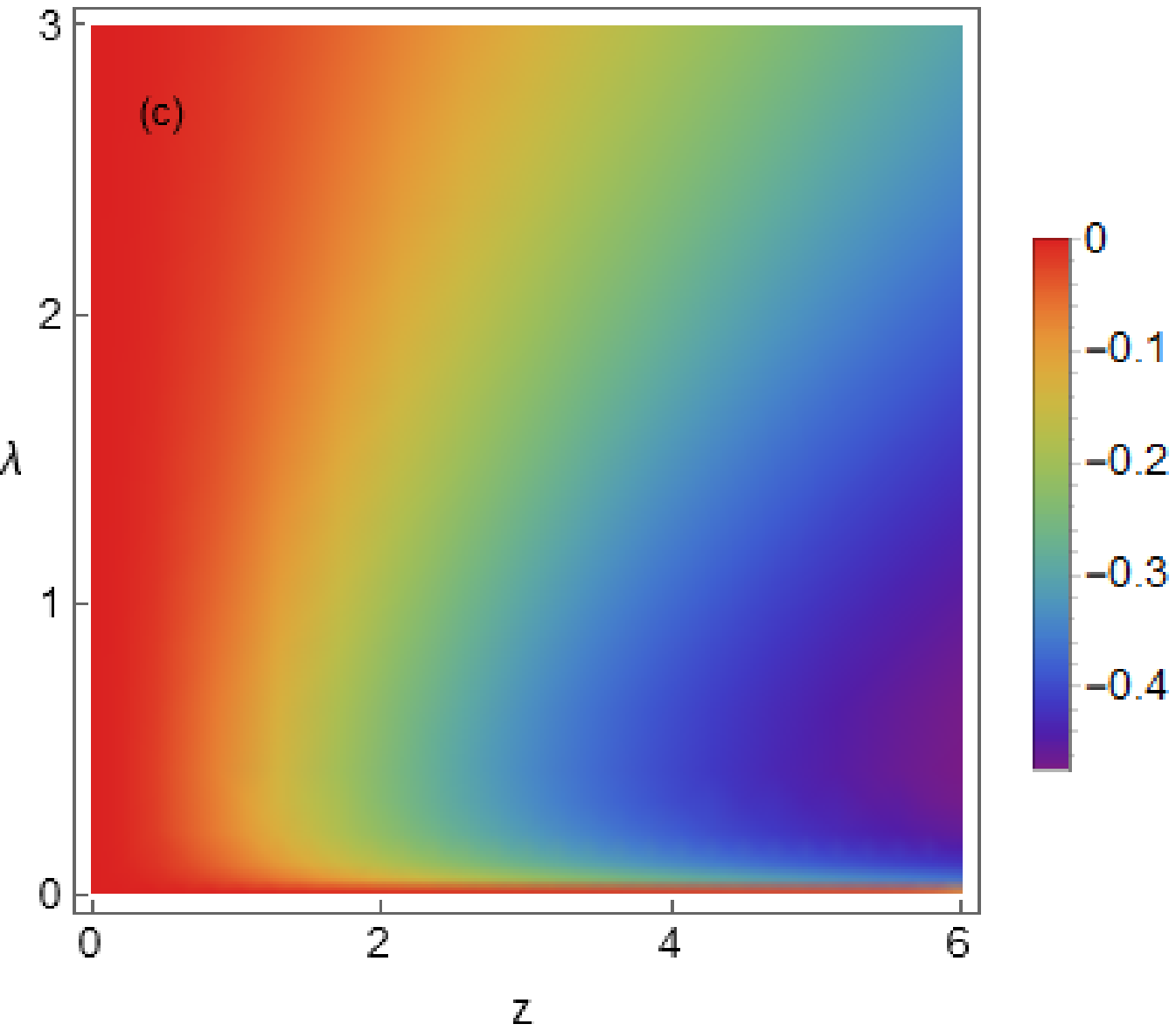}
\caption{The Mandel parameter of the NCSsOC as a function of $\lambda$ and $z$ (a,b). (c) The density plot of Mandel parameter for the NCSsOC versus $z$ and $\lambda$.}
\label{fig:Mandel}
\end{figure}
In this subsection, we investigate deviation from the Poisson distribution for the NCSsOC using the Mandel parameter. This parameter is given by~\cite{Wolf}:
\begin{equation}\label{Mandel}
Q= \frac{{{{(\Delta n)}^2} - \left\langle n \right\rangle }}{{\left\langle n \right\rangle }},
\end{equation}
where the positive, zero, and negative values represent super-Poissonian, Poissonian, and sub-Poissonian distribution, respectively. Due to the complexity of the final form of the Mandel parameter for the NCSsOC, we do not attempt to obtain the analytic form. Instead, we numerically investigate the Mandel parameter for such states using Eqs.~(\ref{mean number of photons}) and~(\ref{probability of finding n photon}).
In Fig.~\ref{fig:Mandel}, we have plotted the Mandel parameter of the NCSsOC as a function of $\lambda$ and $z$.
The results show that, for some fixed values of $z=0.1, 1, 2\,$ and $3$, the Mandel parameter of the NCSsOC first becomes more sub-Poissonian with increasing $\lambda$ and then tends to zero at large values of $\lambda$. Furthermore, this parameter is more negative with increasing $z$ (see Fig.~\ref{fig:Mandel} (a)). Whereas, for some fixed values of $\lambda=0.01, 1, 2 \,$ and $3$, the Mandel parameter becomes more negative with increasing $z$ and then tends to $-0.6$ ($-1$) at large (small) values of $\lambda$, as seen in Fig.~\ref{fig:Mandel} (b). We can justify this result by the fact that the NCSsOC behaves as the standard CSs at small $\lambda$ and when the parameter $z$ is large enough. Indeed, the NCSsOC is the statndard CSs at exactly zero curvature, as shown in Eq.~(\ref{CSs}).
In Fig.~\ref{fig:Mandel} (c), we observe that at the small value of $\lambda$ the most negative amount of the Mandel parameter occurs in the NCSsOC with the small $z$ values.
\subsection{Quadrature squeezing}
\begin{figure}[t!]
\centering
\includegraphics[scale=0.4]{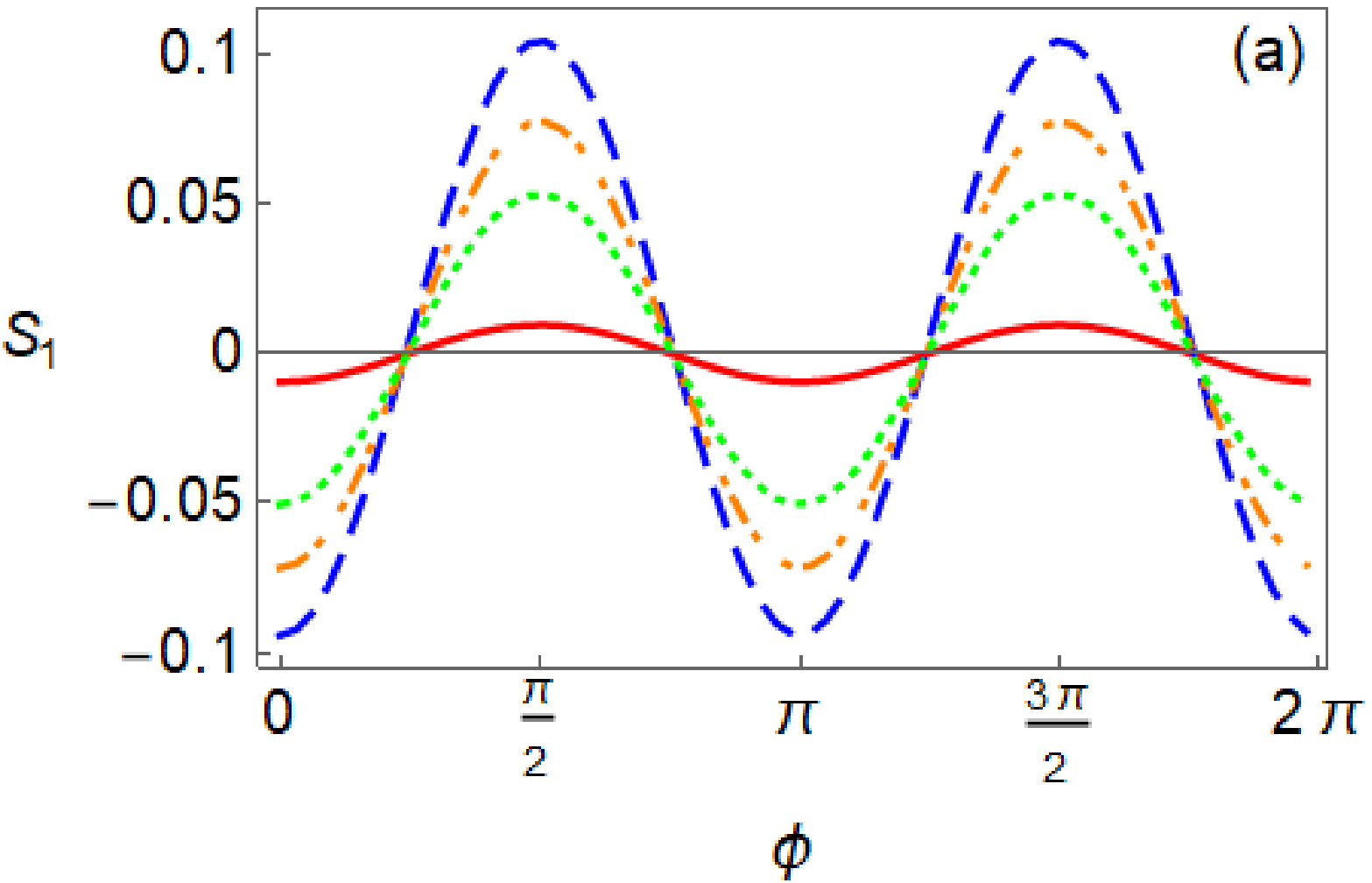}
\includegraphics[scale=0.4]{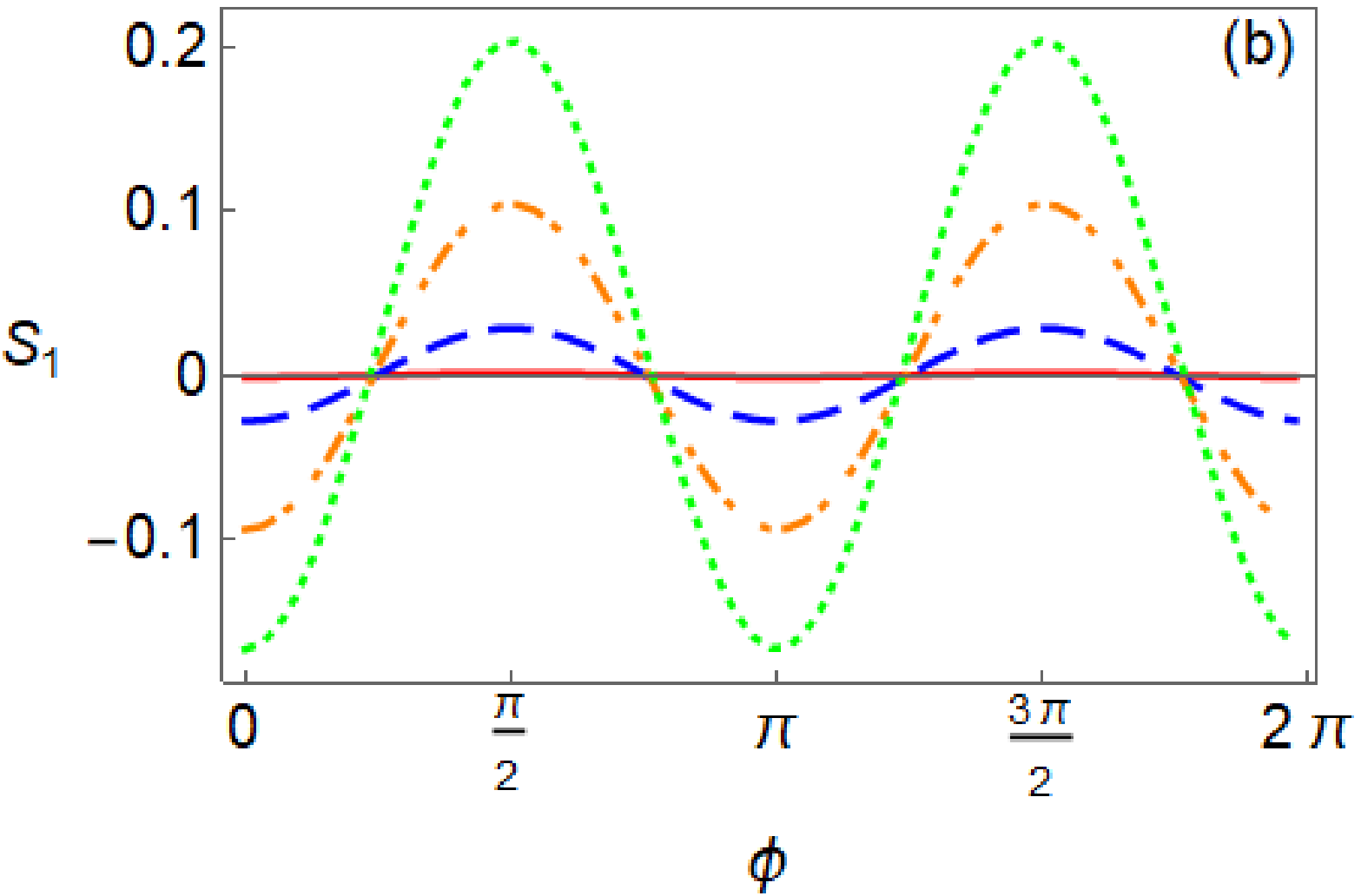}
\includegraphics[scale=0.35]{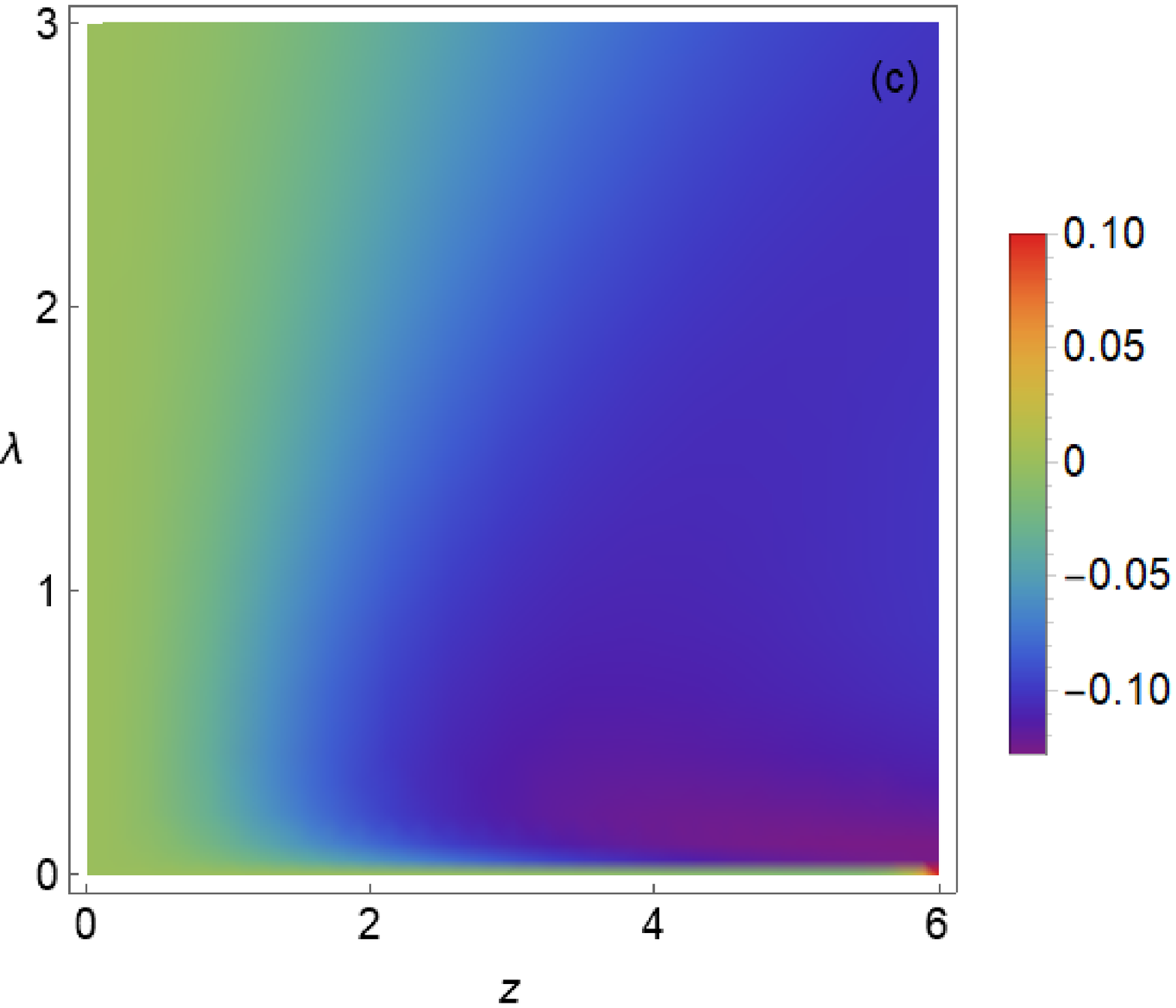}
\includegraphics[scale=0.35]{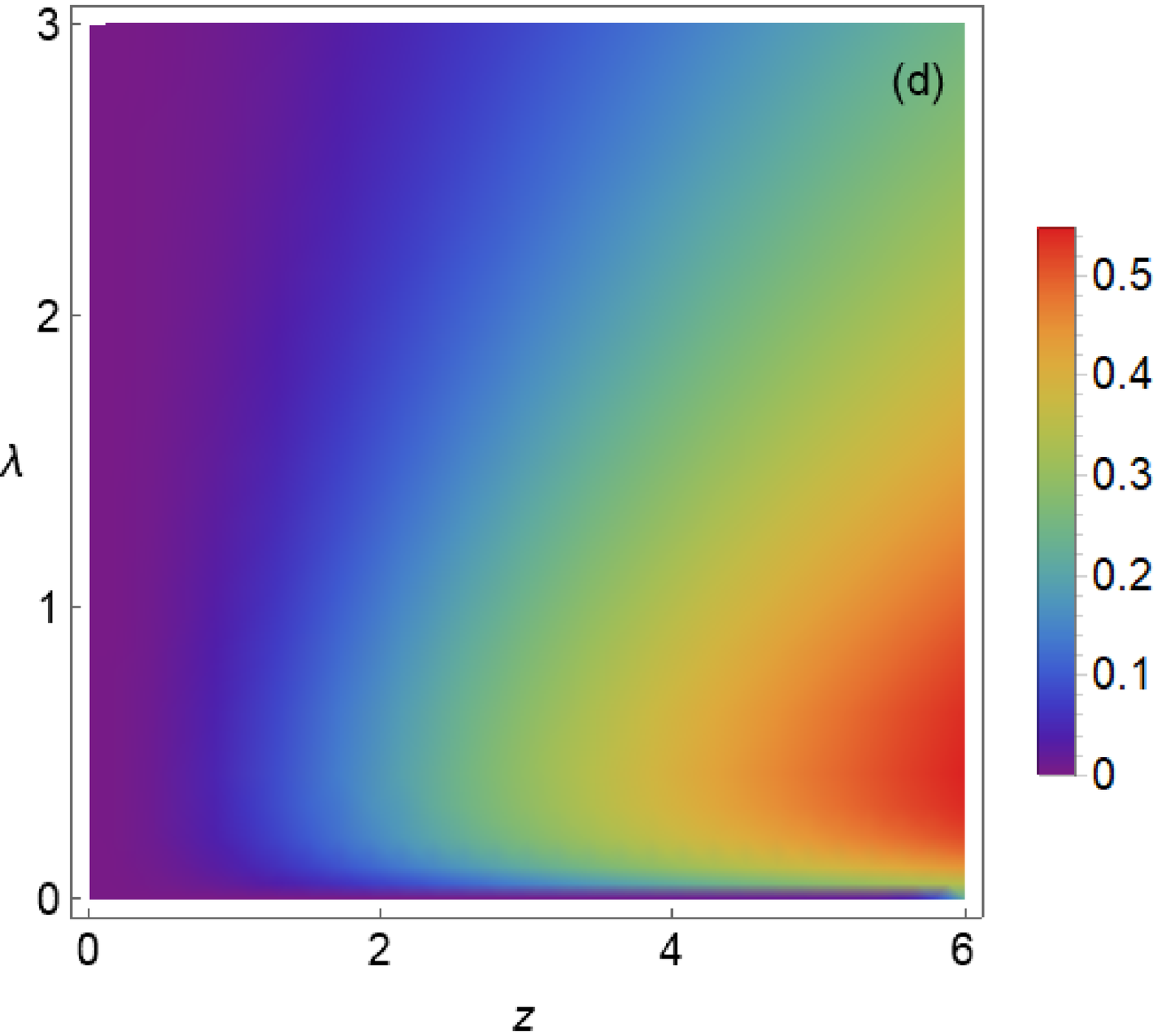}
\caption{The squeezing parameter $S_{1}$ versus $\phi$ for (a) $z=1$, and (b) $\lambda=0.5 $. Here, solid red,
dashed blue, dashdotted orange and dotted green lines correspond to (a) ((b)) $\lambda=0.01$ $(z=0.1)$, $\lambda=0.5$ $(z=0.5)$, $\lambda=1$ $(z=1)$ and $\lambda=1.5$ $(z=1.5)$, respectively. The density plots of (c) $S_1$, and (d) $S_2$ as functions of $z$ and $\lambda$ for $\phi=\pi/6$. }
\label{fig:squeezing}
\end{figure}
In this subsection, we consider  the quadrature operators $\hat X_1$ and $\hat X_2$ defined in terms of creation and
annihilation operators  $\hat a$ and $\hat a^\dagger$ as follows :
 \begin{equation}\label{X1 and X2}
   \begin{array}{l}
    {{\hat X}_1} = \frac{1}{2}\left( {\hat a{e^{i\phi }} + {{\hat a}^\dag }{e^{ - i\phi }}} \right),\\
    {{\hat X}_2} = \frac{1}{2i}\left( {\hat a{e^{i\phi }} - {{\hat a}^\dag }{e^{ - i\phi }}} \right).
   \end{array}
 \end{equation}
By using the commutation relation of $\hat a$ and $\hat a^\dagger$, the following uncertainty relation is obtained
\begin{equation}
{\left( {\Delta {X_1}} \right)^2}{\left( {\Delta {X_2}} \right)^2} \ge \frac{1}{{16}}{\left| {\left\langle {\left[
{{{\hat X}_1},\left. {{{\hat X}_2}} \right]} \right.} \right\rangle } \right|^2} = \frac{1}{{16}}.
\end{equation}
As is known, the quadrature squeezing occurs if we have $ (\Delta X_i)^2<1/4 (i=1 or 2) $ or equally $S_i\equiv
4{(\Delta X_i)^2}-1<0 $. Employing Eqs.~(\ref{mean number of photons}) and~(\ref{X1 and X2}), after some algebra, the squeezing parameter $S_1$ is obtained as
\begin{eqnarray}\label{squeezing S}
S_1&=&\frac{ 2 \cos2\phi}{z^2 \mathcal{N}(z^2)}\sum_{n=2}^\infty \frac{z^{2n}\sqrt{n(n-1)}}{\sqrt{\rho(n-2)\rho(n)}}-\Bigg(\frac{ 2\cos\phi  }{z \mathcal{N}(z^2)}\sum_{n=1}^\infty \frac{z^{2n}\sqrt{n}}{\sqrt{\rho(n-1)\rho(n)}}\Bigg)^2 \nonumber\\
&&+2\langle \hat{n} \rangle.
\end{eqnarray}
Here, $z$ is assumed to be a real parameter, and the mean number of photons, $\langle \hat{n} \rangle$, is already defined in Eq.~(\ref{mean number of photons}). Figs.~\ref{fig:squeezing}(a) and~\ref{fig:squeezing}(b) display the squeezing parameter $S_{1}$ for the NCSsOC as a function of $\varphi$ for $z=1$ and $\lambda=0.5$, respectively. The
red solid, dashed blue, dashdotted orange and dotted green lines depict numerical results related to $\lambda=0.01$, $\lambda=0.5$, $\lambda=1$ and $\lambda=1.5$ for Fig.~\ref{fig:squeezing}(a), and $z=0.1$, $z=0.5$, $z=1$ and $z=1.5$ for Fig.~\ref{fig:squeezing}(b), respectively.
These figures show a periodic dependant on $\phi$, and in areas that the NCSsOC is squeezed the amount of squeezing always enhances with increasing $z$ (see Fig.~\ref{fig:squeezing}(b)). While, for the fixed value $z=1$, the squeezing first enhances with increasing $\lambda$ and then leads to decreasing squeezing, as seen in Fig.~\ref{fig:squeezing}(a)). This behavior can be easily understood from Fig.~\ref{fig:squeezing}(c) that at the small value of $\lambda$ the highest amount of the squeezing appears in the NCSsOC with small $z$, which is also in agreement with the results of the Mandel parameter. At $\lambda=0$, the squeezing properly vanished, as expected from the standard CSs. It is worth noting that the squeezing parameter $S_2$ can be obtained from Eq.~(\ref{squeezing S}) by replacing $\phi=\phi+\pi/2$. In this manner, the squeezing parameter $S_2$ is plotted for different values of $\lambda$ and $z$ in Fig.~\ref{fig:squeezing}(d). This plot clearly exhibits the same behavior of the squeezing parameter $S_1$ with $\phi=5\pi/6$ (not shown here).
\section{Summary and Concluding Remarks}\label{SUM}

In this paper,  we have obtained the Hamiltonian of a harmonic oscillator confined on a circle using the gnomonic projection. It is shown that the algebra of such harmonic oscillator can be regarded as $f$-deformed harmonic oscillator algebra. We have constructed the nonlinear CSs for such harmonic oscillator so that the NCSsOC properly tends to the standard CSs at zero curvature. We then studied the quantum statistical properties of the NCSsOC, and found that the nonclassical features of these states enhance with increasing the parameter $z$ even in small curvatures of the circle.
\\

\section{Acknowledgment}
A.M. wishes to thank The Office of Graduate Studies and Research Vice President of The
University of Isfahan for their support. E.A. also wishes to thank the Shahrekord University for their support.
 
 \vspace{1cm}


\end{document}